\input epsf


\input harvmac.tex
\input tables.tex
%
%
\def\cpt{\vbox{\sl\centerline{Centre for Particle Theory, Department
of Mathematical Sciences}%
\centerline{South Road, Durham, DH1 3LE}}}

\def\sn {{\rm sn}}

\ifx\epsfbox\UnDeFiNeD\message{(NO epsf.tex, FIGURES WILL BE IGNORED)}
\def\figin#1{\vskip2in}
\else\message{(FIGURES WILL BE INCLUDED)}\def\figin#1{#1}\fi
\def\ifig#1#2#3{\xdef#1{fig.~\the\figno}
\goodbreak\midinsert\figin{\centerline{#3}}%
\smallskip\centerline{\vbox{\baselineskip12pt
\advance\hsize by -1truein\noindent\footnotefont{\bf Fig.~\the\figno:} #2}}
\bigskip\endinsert\global\advance\figno by1}
 
\def\bo{ { \sqcup\llap{ $\sqcap$} } }
\overfullrule=0pt       

\def\book#1[[#2]]{{\it#1\/} (#2).}
 
\def\annph#1 #2 #3.{{\it Ann.\ Phys.\ (N.\thinspace Y.) \bf#1} #2 (#3).}
\def\apj#1 #2 #3.{{\it Ap.\ J.\ \bf#1} #2 (#3).}
\def\cmp#1 #2 #3.{{\it Commun.\ Math.\ Phys.\ \bf#1} #2 (#3).}
\def\cqg#1 #2 #3.{{\it Class.\ Quantum Grav.\ \bf#1} #2 (#3).}
\def\foundph#1 #2 #3.{{\it Found.\ Phys.\ \bf#1} #2 (#3).}
\def\grg#1 #2 #3.{{\it Gen.\ Rel.\ Grav.\ \bf#1} #2 (#3).}
\def\intjmp#1 #2 #3.{{\it Int.\ J.\ Mod.\ Phys.\ \rm A\bf#1} #2 (#3).}
\def\intjmpd#1 #2 #3.{{\it Int.\ J.\ Mod.\ Phys.\ \rm D\bf#1} #2 (#3).}
\def\jmp#1 #2 #3.{{\it J.\ Math.\ Phys.\ \bf#1} #2 (#3).}
\def\jphysa#1 #2 #3.{{\it J.\ Phys.\ \rm A\bf#1} #2 (#3).}
\def\mpla#1 #2 #3.{{\it Mod.\ Phys.\ Lett.\ \rm A\bf#1} #2 (#3).}
\def\mnras#1 #2 #3.{{\it Mon.\ Not.\ R.\ Ast.\ Soc.\ \bf#1} #2 (#3).}
\def\nat#1 #2 #3.{{\it Nature\ \bf#1} #2 (#3).}
\def\ncim#1 #2 #3.{{\it Nuovo Cim.\ \bf#1\/} #2 (#3).}
\def\ncimb#1 #2 #3.{{\it Nuovo Cim.\ \bf#1\/}B #2 (#3).}
\def\np#1 #2 #3.{{\it Nucl.\ Phys.\ \bf#1} #2 (#3).}
\def\npb#1 #2 #3.{{\it Nucl.\ Phys.\ \rm B\bf#1} #2 (#3).}
\def\phrep#1 #2 #3.{{\it Phys.\ Rep.\ \bf#1} #2 (#3).}
\def\pl#1 #2 #3.{{\it Phys.\ Lett.\ \bf#1} #2 (#3).}
\def\pla#1 #2 #3.{{\it Phys.\ Lett.\ \bf#1\/}A #2 (#3).}
\def\plb#1 #2 #3.{{\it Phys.\ Lett.\ \bf#1\/}B #2 (#3).}
\def\pr#1 #2 #3.{{\it Phys.\ Rev.\ \bf#1} #2 (#3).}
\def\prb#1 #2 #3.{{\it Phys.\ Rev.\ \rm B\bf#1} #2 (#3).}
\def\prd#1 #2 #3.{{\it Phys.\ Rev.\ \rm D\bf#1} #2 (#3).}
\def\prl#1 #2 #3.{{\it Phys.\ Rev.\ Lett.\ \bf#1} #2 (#3).}
\def\pcps#1 #2 #3.{{\it Proc.\ Cambs.\ Phil.\ Soc.\ \bf#1} #2 (#3).}
\def\plms#1 #2 #3.{{\it Proc.\ Lond.\ Math.\ Soc.\ \bf#1} #2 (#3).}
\def\prs#1 #2 #3.{{\it Proc.\ Roy.\ Soc.\ \rm A\bf#1} #2 (#3).}
\def\revmod#1 #2 #3.{{\it Rev.\ Mod.\ Phys.\ \bf#1} #2 (#3).}
\def\rprog#1 #2 #3.{{\it Rep.\ Prog.\ Phys.\ \bf#1} #2 (#3).}
\def\sovpu#1 #2 #3.{{\it Sov.\ Phys.\ Usp.\ \bf#1} #2 (#3).}
\def\sovjpn#1 #2 #3.{{\it Sov.\ J.\ Part.\ Nucl.\ \bf#1} #2 (#3).}
\def\sovj#1 #2 #3.{{\it Sov.\ J.\ Nucl.\ Phys.\ \bf#1} #2 (#3).}
\def\jtpl#1 #2 #3.{{\it Sov.\ Phys.\ JETP\ Lett.\ \bf#1} #2 (#3).}
\def\jtp#1 #2 #3.{{\it Sov.\ Phys.\ JETP\ \bf#1} #2 (#3).}
\def\zphys#1 #2 #3.{{\it Zeit.\ Phys.\ \bf#1} #2 (#3).}
 
%
%
\lref\KE{S.Kolitch and D.Eardley, \annph 241 128 1995.}
\lref\CCOS{C.Romero and A.Barros, \grg 25 491 1993. \hfill\break
S.Kolitch, \annph 246 121 1996.}
\lref\BDCOS{J.D.Barrow and K.Maeda, \npb 341 294 1990.\hfill\break
A.Burd and A.Coley, \plb 267 330 1991.\hfill\break
J.D.Barrow, \prd 47 5329 1993. \prd 48 3592 1993.\hfill\break
A.Serna and J.Alimi, [astro-ph/9510139], \prd 53 3074 1996.\hfill\break
J.Barrow and P.Parsons, {\it The behaviour of cosmological models
with varying G}, gr-qc/9607072}
\lref\DaN{T.Damour and K.Nordtvedt, \prl 70 2217 1993. \prd 48 3436 1993.}
\lref\HKVW{R.Holman, E.W.Kolb, S.Vadas and Y.Wang, \plb 250 24 1990.}
\lref\GRL{A.Green and A.Liddle, {\it Open inflationary
universes in the induced gravity theory}, astro-ph/9607166}
\lref\GUTH{A.Guth, \prd 23 347 1981.}
\lref\STRING{A.A.Tseytlin and C.Vafa, [hep-th/9109048], 
\npb 372 443 1992.\hfill\break
D.Goldwirth and M.Perry, [hep-th/9308023], \prd 49 5019 1994.\hfill\break
E.Copeland, A.Lahiri and D.Wands, [hep-th/9406216], 
\prd 50 4868 1994.\hfill\break
G.Veneziano, {\it String Cosmology: Basic ideas and general results},
hep-th/9510027 and references therein.}
\lref\CFMP{E.Fradkin, \plb 158 316 1985.\hfill\break
C.Callan, D.Friedan, E.Martinec and M.Perry, \npb 262 593 1985.\hfill\break
C.Lovelace, \npb 273 413 1985.}
\lref\GL{A.Green and A.Liddle, [astro-ph/9604001], \prd 54 2557 1996.}
\lref\RWILL{C.M.Will, \book Theory and Experiment in Gravitational Physics 
[[C.U.P., 1981]]\hfill\break
R.D.Reasenberg et.al., \apj 234 L219 1979.}
\lref\EJW{E.J.Weinberg, \prd 40 3950 1989.}
\lref\ALT{D.La, P.Steinhardt and E.Bertschinger, \plb 231 231 1989.\hfill\break
P.Steinhardt and F.Accetta, \prl 64 2740 1990.\hfill\break
F.Acceta and J.Trester, \prd 39 2854 1989.}
\lref\JBD{P.Jordan, \zphys 157 112 1959.\hfill\break
C.Brans and R.H.Dicke, \pr 124 925 1961.}
\lref\LS{D.La and P.Steinhardt, \prl 62 376 1989.}

\Title{ \vbox{\baselineskip12pt\hbox{DTP-96-55}\hbox{gr-qc/9611065}}}
{\vbox{
\centerline{Cosmology in Brans-Dicke theory with a scalar potential} }} 
\centerline{Caroline Santos\footnote{$^*$}{On leave from: 
Departmento de F\'\i sica da Faculdade de Ci\^encias da Universidade do Porto,
Rua do Campo Alegre 687, 4150-Porto, Portugal} and Ruth Gregory}
\medskip
\cpt
\bigskip

\centerline{ ABSTRACT }
\bigskip

\noblackbox
\noindent We consider the general behaviour of cosmologies in
Brans-Dicke theory where the dilaton is self-interacting via a
potential $V(\Phi)$. We show that the general radiation universe
is a two-dimensional dynamical system whereas the dust or false
vacuum universe is three-dimensional. This is in contrast to the
non-interacting dilaton which has uniformly a two-dimensional
phase space. We  find the phase spaces in each case and the general
behaviour of the cosmologies.
 
\Date{November, 1996}
 
\newsec{Introduction}

Einstein's theory of General Relativity is extremely successful at describing
the dynamics of our solar system, and indeed the observable universe, 
nonetheless, the realization that 
general relativity probably does not describe gravity 
accurately at all scales has led to various alternatives being explored,
most notably the scalar-tensor family of gravity theories pioneered by
Jordan, Brans and Dicke (JBD) \JBD, and the gravitational
lagrangian inspired by low energy string theory, \CFMP.
In fact, these two theories are not unrelated, since the
low energy effective action for bosonic string theory generically
takes the form
\eqn\stract{
S = \int d^4 x \sqrt{-g} e^{-2\phi} \left ( R + 4(\nabla\phi)^2 -
{1\over 12} H_{\mu\nu\lambda}^2 \right )
}
where $\phi$ is the dilaton, and $H_{\mu\nu\lambda}$ is the field
strength of the two form $B_{\mu\nu}$; comparing this with the JBD action
\eqn\bdact{
S_{JBD} = \int d^4x \sqrt{-g} \left [ \Phi R - \omega {(\nabla\Phi)^2\over\Phi}
+ 16\pi {\cal L} \right ]
}
shows that the scalar-tensor sectors are identical if $\omega = -1$.

The phenomenological importance of such extended gravity theories to
modern cosmology was powerfully highlighted by
La and Steinhardt \LS, who suggested that by
using JBD theory instead of general relativity, the ``graceful exit''
problem of old inflationary cosmology \GUTH\  might be ameliorated.
It was rapidly realised \EJW\  that this original scenario required a 
value of the JBD parameter $\omega$ which was in conflict with the 
observational limits \RWILL. This then led to the development of 
various alternatives \ALT, however, a general study of the extended
inflationary ethos \GL\ indicates that in order to satisfy observational
constraints rather contrived models are required.
Nonetheless, extended inflationary ideas still survive in some recent
inflationary models (e.g.\ \GRL). 

The prominence of string theory as a theory of everything, in 
particular a quantum theory of gravity, means that we must examine
its consequences in regimes where it departs from 
general relativity, in particular
we expect that the early universe might display ``stringy'' qualities;
\STRING\ gives a selection of articles dealing
with precisely this problem. Correspondingly, by field redefinitions,
one can instead investigate the cosmological implications of 
JBD theories, which several recent authors have done in a
variety of ways \BDCOS, \DaN, \CCOS, \KE.
Most of the recent studies of the JBD theories
have focussed on the qualitative behaviour
of cosmologies in pure JBD theory, or JBD plus a 
cosmological constant\footnote{$^\dagger$}{It should be stressed that the 
`cosmological constant' in this case amounts to setting $R\to R+\Lambda$
in the JBD action.}.
However, since it is generally believed that the
string dilaton must be massive in order to anchor it to the current
observed Newton's constant, it seems likely
that more realistic JBD models
will include a potential for the dilaton:
\eqn\ouract{
S = S_{JBD} - \int d^4x \sqrt{-g} V(\Phi)
}
The specific form of this potential is unknown, but
it seems reasonable that at high temperatures it might
look something like $V_0\Phi^{2n}$. In this paper therefore,
we are interested in the qualitative behaviour of
JBD cosmologies with just such a form for the potential,
focusing in particular on $n=1$: $V(\Phi) = V_0\Phi^2$.
(The cosmological constant models investigated in  \CCOS\ have
$V(\Phi) = \Lambda\Phi$.) Clearly this is not a particularly
realistic potential for the current time, since it has a minimum
at $\Phi=0$, which would correspond to an infinitely strong
gravitational coupling, (although as we will show $V(\Phi)$ is not
necessarily to be viewed as a canonical scalar field potential)
however, the model potential ought to provide
some insight on the cosmological effects of a self-interacting
dilaton.

Our starting point will be the recent papers of Kolitch and
others\KE, who showed that the field equations for cosmological models
could be reduced to a two-dimensional dynamical system
for any reasonable perfect fluid matter source in pure JBD theory,
and also for spatially flat models with a non-zero
cosmological constant.  In the presence of the scalar
potential, the cosmological dynamical system turns out to be 
generically three-dimensional, apart
from radiation dominated universes in which the system once more
reduces to two-dimensions. The layout of the paper is as
follows. In the next section we analyse the cosmological equations
of motion following from \ouract\ and derive the corresponding
minimal dynamical system for a spherically symmetric perfect fluid
cosmology. In the third section we consider the example of empty
universes as a means of testing the method and exploring the
effect of the potential on the purely gravitational theory, we also
consider radiation universes, since these too turn out to have a
particularly simple form qualitatively similar to the vacuum spacetimes.
We also consider conformally transforming to 
the so-called Einstein frame to motivate some of  these
features. In section four we present an analysis of dust and
false vacuum models.  Finally,
we sum up our results and conclude in section five.
We use a mostly plus signature for the metric.

\newsec{The cosmological equations.}
 
In this section we present a brief derivation of the dynamical
system representing a spherically
symmetric cosmology in JBD theory with the scalar potential.
Varying the action \ouract\ with respect to the metric and scalar field yields
the following equations of motion:
\eqn\bdein{
R_{\mu\nu} - {\half} R g_{\mu\nu} -
 {\omega\over\Phi^2} \nabla_\mu \Phi \nabla_\nu\Phi
- {1\over\Phi} \nabla_\mu \nabla_\nu \Phi + g_{\mu\nu}
\left ( {\bo\Phi \over \Phi} + {\omega \over 2\Phi^2} (\nabla\Phi)^2
+ {V(\Phi) \over 2\Phi} \right )
= {8\pi\over\Phi}  T_{\mu\nu}
}
\eqn\phhi{
R - {\omega \over \Phi^2} (\nabla\Phi)^2 + {2\omega \over \Phi} \bo \Phi
- {dV(\Phi)\over d\Phi} =0
}
Contracting \bdein\ and substituting in \phhi\ gives 
\eqn\feom{
\bo \Phi = {8\pi \over 3+2\omega} T + {1\over 3+2\omega} \left (
\Phi {dV(\Phi) \over d\Phi} - 2 V(\Phi) \right )
}
as a reduced equation of motion for $\Phi$. 

Following the cosmological principle, we will assume a standard
Friedman-Robertson-Walker (FRW) form for the metric:
\eqn\frw{
ds^2 = -dt^2 + a^2(t) \left [ {dr^2 \over 1-kr^2} + r^2 (d\theta^2
+ \sin^2\theta d\phi^2) \right ] .
}
Substituting into \bdein\ and \feom\ gives the following equations
\eqn\full{
\eqalign{
{{\ddot a} \over a} + 2 \left ( {{\dot a}\over a} \right )^2 + {2k\over a^2}
+ {{\dot a}{\dot \Phi} \over a\Phi} &= {8\pi \over (3+2\omega)\Phi}
[-\omega p + (\omega+1)\rho] + {V(\Phi) \over 2\Phi} \cr
\left ( {{\dot a}\over a} + {{\dot \Phi} \over 2 \Phi} \right )^2 +
{k\over a^2} &= {2\omega +3 \over12} \left ({{\dot \Phi} \over \Phi} \right )^2
+ {8\pi \rho \over 3\Phi} + {V(\Phi) \over 6\Phi} \cr
- {1\over a^3} {d\over dt} ({\dot\Phi} a^3) &= {8\pi \over 3+2\omega}
\left ( T + {\Phi \over 8\pi} {dV(\Phi) \over d\Phi} - {V(\Phi) \over 4\pi}
\right )\cr}}
for the universe. Note that we have assumed the stress energy
takes the form of a perfect
fluid, $T_{ab} = $diag$(\rho,p,p,p)$ in an orthonormal frame. This
satisfies the conservation equation
\eqn\cons{
{\dot \rho} = -3 {{\dot a}\over a} (\rho +p)
}

Following \KE\ we transform to conformal time 
\eqn\conft{
\eta = \int {dt\over a(t)}
}
and we take an equation of state 
\eqn\eostate{
p = (\gamma-1)\rho
}
for the fluid. Note that this implies
\eqn\rhoofa{
\rho \propto a^{-3\gamma}
}
from \cons.

Denoting ${d\over d\eta}$ by a prime, and defining
\eqn\define{
\eqalign{
X &= \left [{2\omega+3\over12} \right ]^{1/2} 
{\Phi'\over\Phi} = A{\Phi'\over\Phi} \cr
Y &= {a'\over a} + {\Phi'\over2\Phi} \cr
}}
we see that for the potential $V(\Phi) = V_0 \Phi^2$,
the latter two equations of \full\ become:
\eqn\interX{
X' = - 2 XY + (1-3\gamma/4) {8\pi \rho a^2\over 3A\Phi}
}
\eqn\interY{
Y^2 + k = X^2 + {8\pi \rho a^2\over 3\Phi} +
{V_0\Phi a^2 \over 6}  }
we may then, \KE, differentiate this constraint equation for $Y$,
using the $X$-equation and the equation of motion for $\rho$ to finally
obtain
\eqn\ds{
\eqalign{
X' &= -2XY + {(1 - 3\gamma/4)\over A} (Y^2 + k - X^2
- {Z\over6}) \cr
Y' &= (1 - 3\gamma/2) (Y^2 + k - X^2) -
2 X^2  + {\gamma Z \over 4} \cr
Z' &= 2ZY \cr
}}
where $Z = V_0 \Phi a^2$, as the three-dimensional dynamical system
representing the general perfect fluid cosmological models for a JBD
theory with the scalar potential $V_0\Phi^2$. (The method can be
generalised for $V(\Phi) = V_0 \Phi^n$ in which case $Z = V_0 \Phi^{n-1}
a^2$, and $X'$ acquires additional $Z$-terms.)

Note that \ds\ represents the most general case scenario.
If $V_0=0$, the $Z$-equation disappears and we recover the Kolitch and
Eardley scenario \KE. Also, if $\rho=0$ (as we will consider in the
next section) \interY\ can be used to eliminate $Z$, again leading
to a two-dimensional dynamical system, and in fact, we can also reduce \ds\ to a
two-dimensional system in a radiation universe, since $\rho\propto a^{-4}$ which
allows us to eliminate $X$ using \interY.

Before proceeding to an analysis of \ds\ for the various cosmologies, 
we conclude this section with a few general remarks. First, note that in order
for the Ricci term in the action (1.2) 
to have the correct sign, we require
$\Phi>0$, i.e.\ $Z>0$. From \ds\ we note that since $Z'|_{Z=0}
=0$, the dynamical system trajectories will not cross $Z=0$. We will also
take $\omega > -3/2$, as is conventional, which implies that $A>0$; for the
case of string theory,  $A = {1\over2\sqrt{3}}$. For a physical cosmology,
we will require $\rho>0$, i.e.,
\eqn\posen{
Y^2 + k - X^2 - {Z \over6} \geq 0
}
from the constraint equation \interY. Using the dynamical system
equations, \ds, verifies that
$\left [ Y^2 + k - X^2 - {Z \over6} \right ] '
\Big | _{\rho=0} = 0$
and hence trajectories do not cross $\rho=0$. In other words, a positive
energy trajectory will remain a positive energy trajectory. 

Finally, it
will be of interest to note whether the cosmological models corresponding
to the dynamical system trajectories are expanding or contracting. From
\define\ we note that for expansion $Y - X/2A>0$, therefore we will
be interested in where this surface is relative to the physical regions and
the possible trajectories. It will mostly be the case that trajectories
will lie on one side or the other of this surface, and hence will represent
cosmologies that are either eternally expanding or contracting, however, 
particularly for small $A$, it may well be the case that some trajectories
cross this line, in which case they will correspond to cosmological
solutions which `bounce' in the Brans-Dicke frame, that is, they start off
contracting, reach a minimum size and re-expand (or vice-versa). However,
it should be pointed out that universes which bounce in the JBD frame, may
well not be true bounce universes when viewed from the `Einstein' frame,
a conformally related metric in which the gravitational part of
the action appears in Einstein form. For a universe to bounce in the
Einstein frame requires $Y$ to change sign, something which is only 
possible in $k = 1$ cosmologies.

\newsec{Vacuum and radiation cosmologies.}

We consider vacuum cosmologies first, since these should 
display the main features of the effect of the potential, however,
we also include  radiation universes since these too are effectively 
two-dimensional and are exactly soluble. We first derive the 
two-dimensional dynamical system governing the radiation/vacuum
cosmologies, determining the form of their solution, and plotting the
curves on the two-dimensional phase plane. Using also the original
$X,Y,Z$-variables, we identify the various qualitative behaviours
possible, re-interpreting key solutions in terms of the original
cosmological parameters $\Phi(t)$, $a(t)$.

Setting $\gamma = 4/3$ in \ds\ gives
\eqn\radnds{
\eqalign{
X' &= -2XY \cr
Y' &= -(Y^2 + k +X^2) +{Z \over 3} \cr
Z' &= 2ZY \cr
}}
The key feature that enables radiation (and hence also vacuum) universes
to be simplified is that for $\gamma = 4/3$, \rhoofa\ implies
$\rho \propto a^{-4}$, which means that the constraint equation
\interY\ reads
\eqn\vrcon{
Y^2 + k = {8\pi\rho_0 V_0 \over 3Z} + X^2
+ {Z\over6}
}
This in turn allows us to decouple $X$ from \radnds\ leading to
\eqn\intmd{
Y' = -2(Y^2+k) + {8\pi \rho_0V_0 \over 3Z} + {Z\over2}
}
In order to reduce this to a standard form, we write
\eqn\wdef{
W=YZ\;\;\; (=Z'/2)
}
and obtain
\eqn\radtwo{
\eqalign{
Z' &= 2 W \cr
W'&= {Z^2\over2} - 2kZ + {B\over2} \cr
}}
where we have written $B=16\pi\rho_0V_0/3$ for convenience.

It is not difficult to see that
\eqn\wofz{
W^2 = {Z^3\over 6} - kZ^2 +{BZ\over2} +C
}
satisfies \radtwo, where $C$ is a non-negative constant from \vrcon.
Since $W=Z'/2$, \wofz\ is in fact an elliptic equation, and hence
$Z(\eta)$ can be written in closed form:
%
\eqn\exactz{
Z = Z_1 + {Z_3 - Z_1 \over \sn^2 \left [ - {\sqrt{Z_3-Z_1\over6}}\eta,
\sqrt{Z_2-Z_1\over Z_3-Z_1} \right ]}
}
where the $Z_i$ are the roots of the cubic $W^2(Z)$.
However, since this leads to rather involved general expressions for 
$\Phi$ and $a$ in terms of integrals of elliptic functions, it proves to be
more illuminating to proceed with the qualitative
picture.

First we identify the critical points of the $(Z,W)$-system,
i.e.\ points at which $W'=Z'=0$ which represent an equilibrium solution
of this system of equations. 
From \radtwo, \wofz, these are
\eqn\radcrit{
\eqalign{
P_+ &= (2 + \sqrt{4 - B},0) \; ; \;\;{\rm with\ }\; C = C_0(B), k = 1 \cr
P_{0,-} &= (0,0) \; ; \;\; B=C=0, k = 0,-1\cr
}}
We will return to the
classification of these critical points later.

Next, note that if $W^2(Z)$ is always positive then $W$
is always defined. This is clearly true for all $B$ and $C$ when
$k = 0, -1$, and for $B\geq 3$ if $k = 1$. If, however, $W^2(Z)<0$
at any point, then $W$ cannot be defined and the corresponding
trajectory will consist of disjoint segments. This means that 
there are three main forms that the $(Z,W)$ phase plane can take, 
which are illustrated in figure 1, according to whether the cubic
$W^2(Z)$ has three, two or one real root(s) for the $C=0$ trajectory.
(Figure 1d is the special case $B=0$ of $k=1$.)
Although these figures are shown for $k=1$,
the diagrams are qualitatively the same for $k = 0,-1$ with these 
cases generically taking the qualitative form of figure 1c, although
for $B=0$ (vacuum) the figure includes a critical point at the origin. 
Not surprisingly 
the three different pictures translate into three (slightly) different
qualitative behaviours for the cosmologies.
\ifig\figone{The (Z,W) phase plane for the radiation cosmologies. Critical
points are indicated by a disc, the disallowed $C<0$ and $Z<0$ regions 
are shaded out. The plots are shown for $k = 1$ cosmologies, but the general
form of $k - 0,-1$ cosmologies is qualitatively the same as figure 1c.}
{\epsfxsize 14truecm \epsfbox{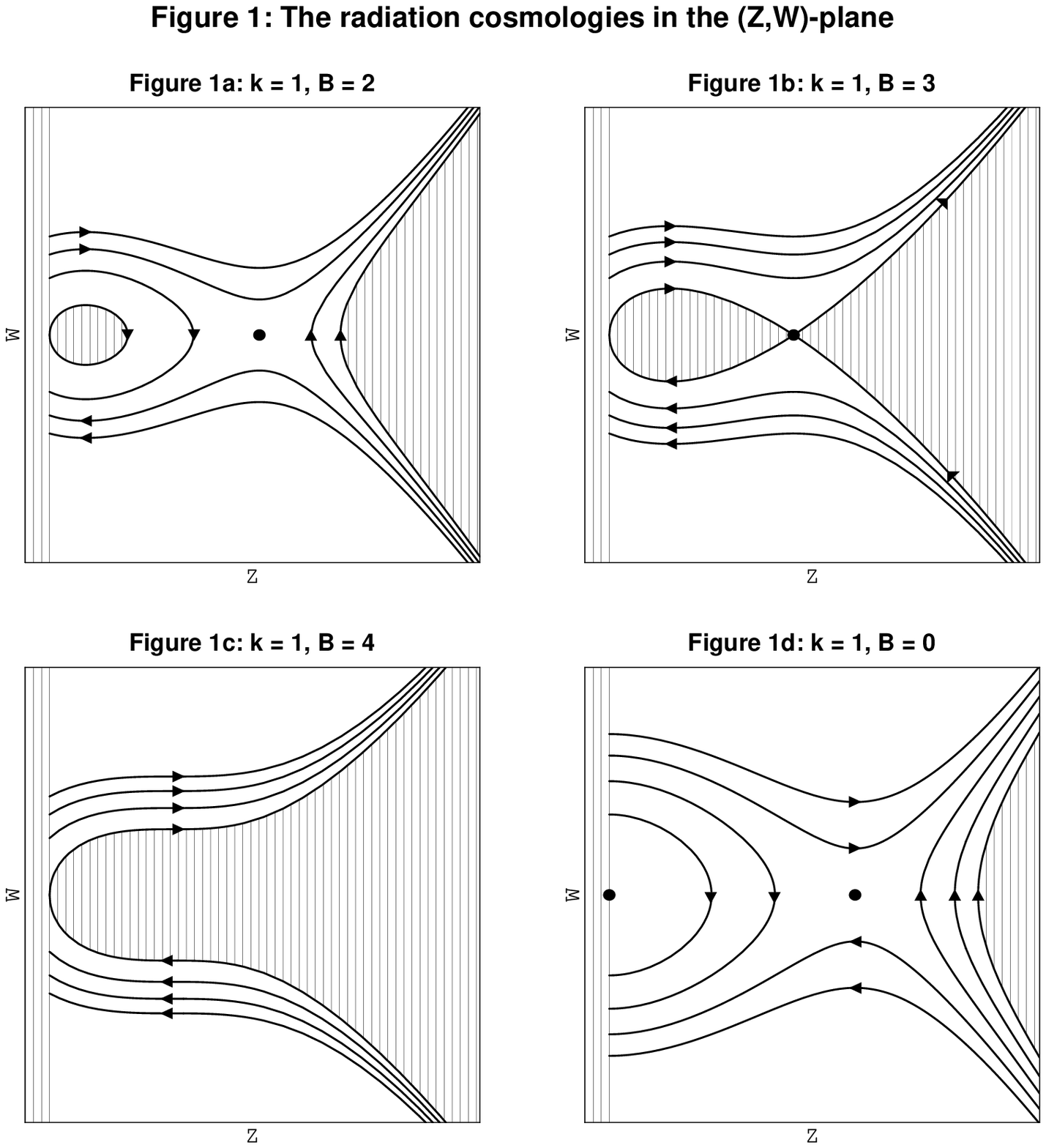}}

Before turning to this however, we first investigate a little further the
parameter ranges for each option. Note that $W^2(Z)$ always
has one non-positive root, $Z_1\leq0$. Whether or not
it has additional real roots depends on whether the discriminant
\eqn\discrim{
D = B^3 - 3B^2k^2 + 18kBC +9C^2-48k^3C
}
is negative or zero.

\noindent $\bullet$ If $D<0$ for $C=0$, then $D<0$ for some range of
$C<C_0$, say, where $C_0$ is the critical value of $C$ for which
$D(B, C_0)=0$:
\eqn\cdefn{
C_0 = {8\over3} k^3 - kB + {1\over3} \left ( 4k^2 - B\right )^{3/2}
}
Therefore, for $C<C_0$,  $W^2(Z)$ has two
additional positive roots for some range of $C$ and we have diagram
1a. This will be the case for $k = 1, B<3, C<C_0$. 
A graph of $C_0(B)$ is shown in figure 2.
\ifig\figtwo{A plot of the parameter region in $(B,C)$ space in which
the discriminant, $D$, is negative. When $D$ can be negative the 
$(Z,W)$ phase plane takes the form of figure 1a and we have
bounce cosmologies. This can only happen for $k = 1$.}
{\epsfxsize 14truecm \epsfbox{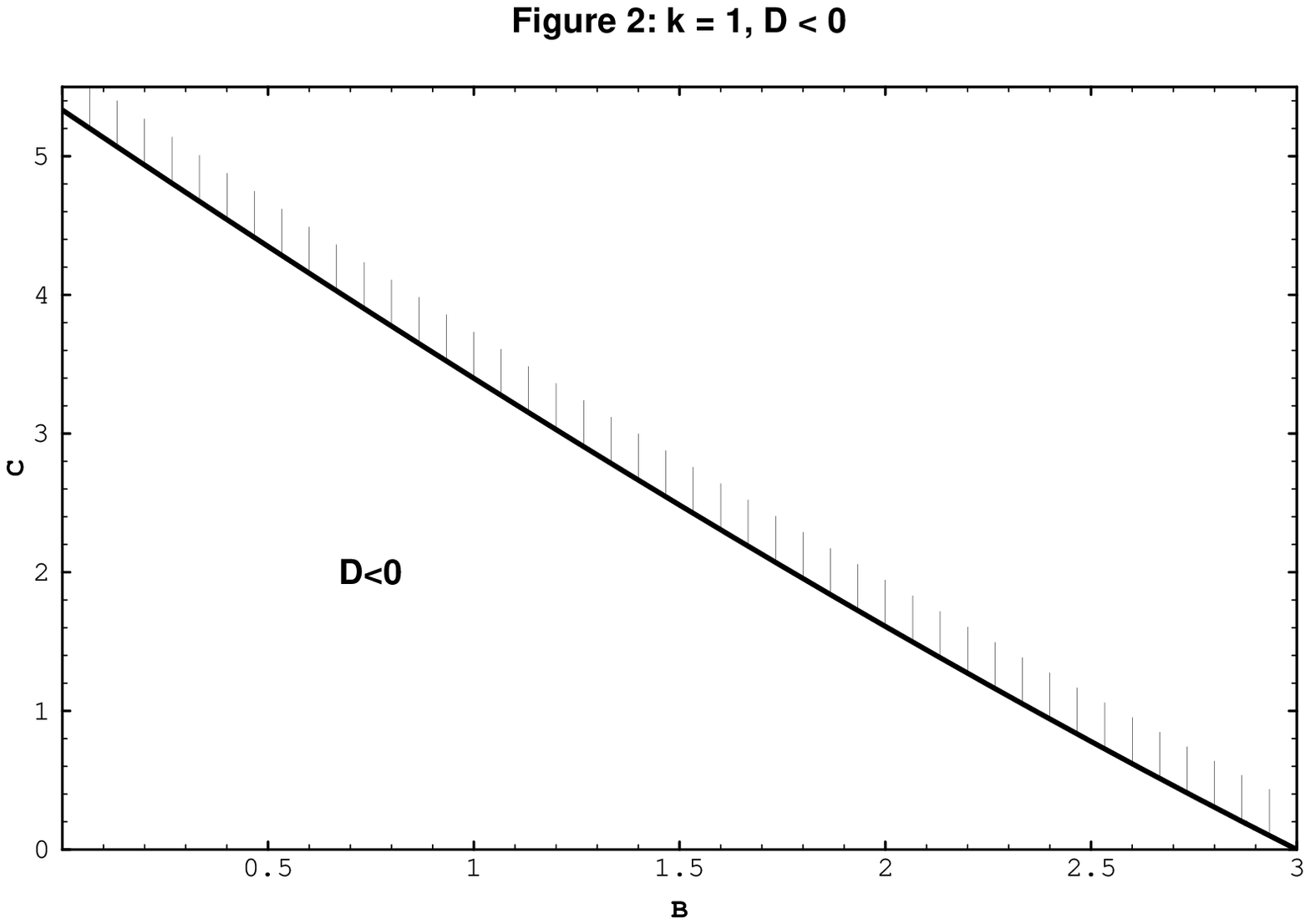}}

\noindent $\bullet$ If $D=0$ for $C=0$ then $W^2(Z)$ has one
(repeated) root for $C=0$. If $k = 1$, $B = 0$ then $D$ is negative on
$(0,C_0)$ and we have diagram 1d, otherwise all other trajectories have no
roots and are continuous. Examining \discrim\ shows that
$B = 3k^2,0$ for $D=C=0$, and the repeated root is therefore
\eqn\rproot{
Z_+ = 3k \;\; ;\;\;\; Z_0=0
}
according to whether $B = 3k^2$ or $B = 0$. If $k = 1$, $B=3$,
$Z_+=3$, and we have diagram 1b. 
If $k=0,-1$, we can only have $B=0$, and we get 
diagram 1c (with the critical point at the origin
as already mentioned).

\noindent $\bullet$ If $D>0$ for $C=0$ then $W^2(Z)$ has no
zeros and we have diagram 1c. This will be the case for all
positive $B$ for $k = 0,-1$, and for $B>3$ if $k=1$.

In terms of interpreting cosmological solutions it is easier to examine
the original $X,Y$ variables. Using \vrcon, \wdef, and \wofz\ it is
easy to see that
\eqn\xyofz{
\eqalign{
X^2 &= {C\over Z^2} \cr
Y^2 &= {Z\over 6} - k + {B\over 2Z} + {C\over Z^2}\cr
}}
which can be used to produce parametric plots of the $(X,Y)$ plane. Although 
these are broadly similar for radiation and vacuum
spacetimes, the vacuum plot is a true two-dimensional dynamical system
whereas once the energy density of radiation is nonzero, the
plot becomes a projection and contains apparent crossings of trajectories. 

In translating from the $(Z,W)$ plane to the $(X,Y)$ plane there will be,
broadly speaking, three qualitatively different pictures.

\noindent$\bullet$ Figure 1a contains
trajectories along which $D<0$, and hence $Y$ has a zero. If we order the roots
of $W^2(Z)$ as $Z_1<Z_2\leq Z_3$, then these trajectories have two branches:
the $Z\in[0,Z_2]$ branch and $Z>Z_3$ branch. For $Z>Z_3$ the trajectory starts
at $(0,-\infty)$ with  $Z$ infinite, and crosses the axis at $X = \pm
\sqrt{CA}/Z_3$ where $Z$ is at a minimum, approaching $(0,\infty)$, 
$Z\to\infty$. Since $X$ always has the same sign, $\Phi$ is either
monotonically increasing or decreasing along such trajectories. This
means that it must be $a(\eta)$ that is causing $Z$ to be infinite, reducing to 
a minimum, and increasing again, i.e.\ this is a bounce cosmology both
in the Einstein and the JBD frame.
For the other branch, $Z$ starts at zero, is maximized at
$Z_2$, returning to zero again. The corresponding trajectories start at
$(\pm\infty, \infty)$ and tend to $(\pm\infty,-\infty)$.
In this case, $X$ once again has a fixed sign, however, it is no longer small,
hence $\Phi$ either increases from $0$ to $\infty$ or vice versa. These 
solutions will also be bounce universes in either frame.
For the critical $D=0$ trajectory there are again
two branches, $Z>Z_0$ and $Z<Z_0$, where $Z_0=Z_2=Z_3$ is the repeated
root. Each of these branches corresponds to four trajectories, each of which
interpolates between the images of the critical point $P_+$ in the $(X,Y)$
plane : $({\pm\sqrt{CA}\over Z_0},0)$, and $(0,\pm\infty)$ for $Z>Z_0$, or 
$(\pm\infty, \pm\infty)$ for $Z<Z_0$. For
$D>0$ there are no zeros of $Y$ and trajectories interpolate between
$\{(\pm\infty, \pm\infty),\  Z=0\}$,
and $\{ (0,\pm\infty),\ Z=\infty\}$.

\noindent$\bullet$ Figure 1b has one $D=0$ 
trajectory which terminates at (0,0); all others are of the $D>0$ form.

\noindent$\bullet$ Figure 1c has $D>0$ for all trajectories hence 
no trajectories cross or terminate on the $X$-axis and
the phase plane consists of two disjoint parts
with  a `barrier' across
the plane separating positive and negative $Y$.

Such are the broad features of the phase planes, however, since vacuum
cosmologies are a special case, we analyse them first before proceeding to the
general case.

\subsec{The vacuum cosmologies.}

Although we have solutions for $X$ and $Y$, we first show how  they form a
two-dimensional dynamical system. For $\rho=0$, \interY\ can be
used to eliminate $Z$ from \ds\ leaving 
\eqn\trvacsis{
\eqalign{
X' &= -2XY \cr
Y' &= Y^2 + k - 3X^2 
}}
independent of the value of $A$.
This dynamical system has four possible critical points:
\eqn\trvacpoin{
\eqalign{
S_{0,1} &= (0,\pm\sqrt{-k}) \;\;\leftrightarrow\;\; P_-, P_0;\cr
S_{2,3} &= (\pm\sqrt{k\over3},0) \;\leftrightarrow\;\; P_+, P_0 \cr
}}
depending on $k$.  These points are classified in table 1, and
the phase planes are shown in figure 3. 
\midinsert
\begintable
\multispan{4}\tstrut\hfil {\bf TABLE 1: 
The critical points of a vacuum cosmology.}\hfil\crthick
\tstrut {~} | $ k = -1$ | $k = 0$ | $k = +1$ \crthick
$S_{0,1}$ | $(0,\pm 1)$, SADDLE | (0,0), DEGENERATE | --- \cr
$S_{2,3}$ | --- | (0,0), DEGENERATE | $(\pm1\over\sqrt{3},0)$, SADDLE \endtable
\endinsert
%
\ifig\figthree{The $(X,Y)$ phase plane diagrams for the vacuum
cosmologies. The shaded areas represent the disallowed regions where
$Z<0$. Critical points are indicated with a dot. Figure 3d shows a closeup
of the critical points in figure 3b, the grey line represents $a'=0$
for $A=0.75$.}
{\epsfxsize 14truecm \epsfbox{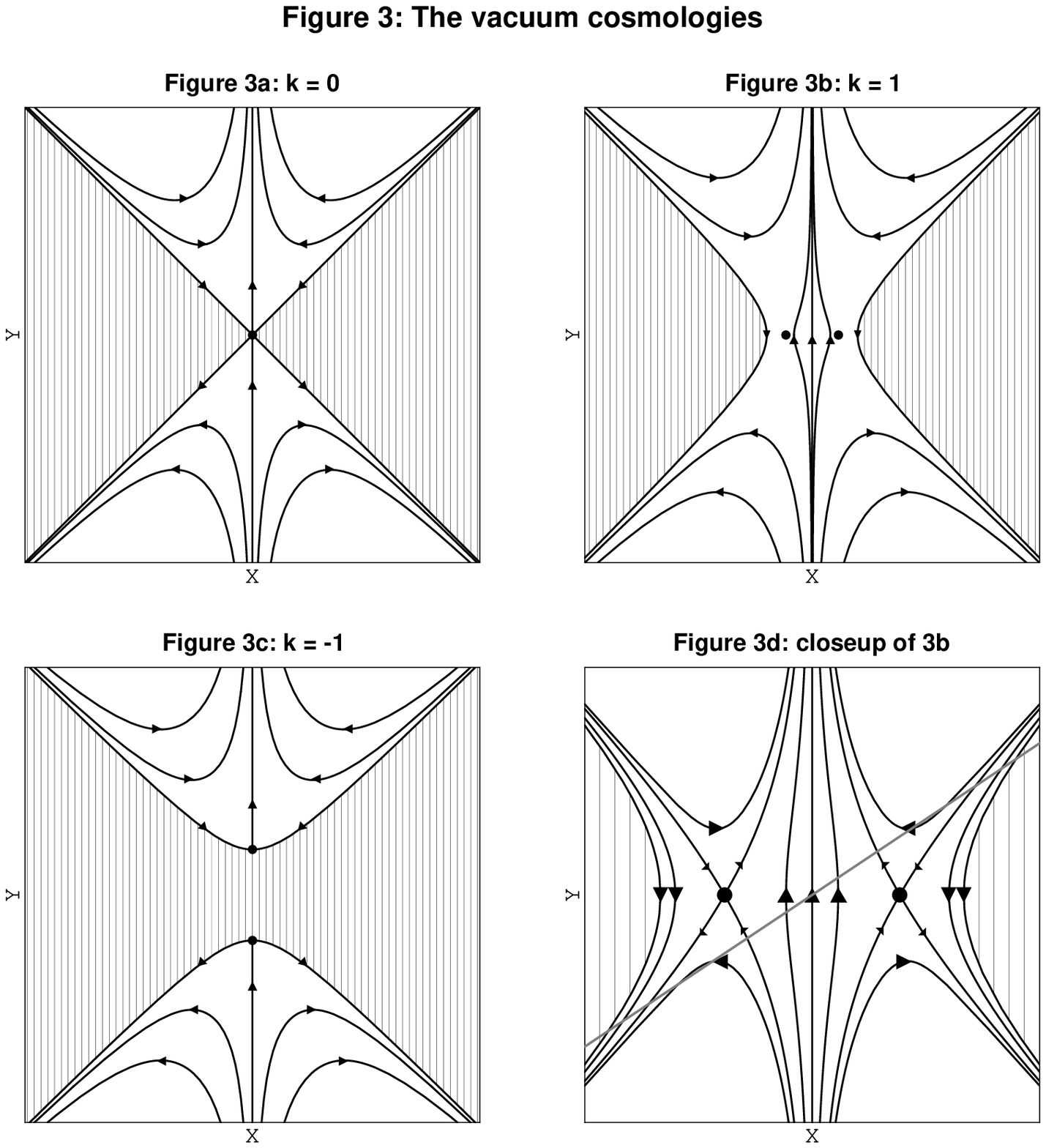}}

In order to interpret the phase diagrams, we must translate from the $X,Y$
variables to a cosmological solution $\Phi(t) , a(t)$. We begin with the 
critical points.

\noindent{\it (i) Critical point solutions.}

\noindent$\bullet$ $k = 0$: Critical point $(0,0)$:
For $X=Y=0$, $a$ is constant and we observe from \interY\ that $\Phi=0$.
Hence this ``cosmology'' is simply Minkowski spacetime.

\noindent$\bullet$ $k = 1$: Critical points $({\pm1\over \sqrt{3}},0)$:
Here the constraint equation \interY\ gives $Z\equiv 4$, a constant.
Solving for $a$ then gives
\eqn\vcritneg{
\eqalign{
a(t) &= {\mp t\over 2\sqrt{3A}} \cr
\Phi(t) &= {48A\over V_0 t^2} \cr
}}
This now has a non-trivial JBD scalar field. The upper (lower) branch
corresponds to the $+ (-)$ critical point, has $t\in(-\infty,0)$
( $t\in(0,\infty)$) and is collapsing to (expanding from) a
singularity.

\noindent$\bullet$ $k = -1$: Critical points $(0,\pm1)$:
Here \interY\ gives $\Phi \equiv 0$ again, however, $a'/a=\pm1$ is
non-zero. Integration gives $a\propto e^{\pm\eta}$ hence
$a = |t|$. The $+ (-)$ root corresponds to $t > (<)0$ and corresponds
to an expanding (contracting) universe, however it should be pointed
out that this is a coordinate transform of Minkowski
spacetime, and is known as the Milne universe.

In addition to the critical point solutions, there are also some
simple solutions to \trvacsis.

\noindent{\it (ii) Constant scalar exact solutions.}

An examination of \trvacsis\ readily reveals that the $Y$-axis
(or segments thereof) represents a solution. Inputting $X=0$ gives
\eqn\desit{
Y(\eta) = \cases{ \tan\eta & $k = 1$\cr
{-1\over\eta} & $k = 0$ \cr
-\coth\eta & $k = -1$ \cr}
}
\noindent$\bullet$ $k = 0$: The $Y$-axis consists of two solution segments,
$Y>0$ and $Y<0$, for which
\eqn\zds{
a(t) = e^{\pm Ht} \;\;\; ; \;\;\; \Phi = {6H^2\over V_0}
}

\noindent$\bullet$ $k=1$: The whole of the $Y$-axis is a trajectory,
which corresponds to the cosmological solution
\eqn\pds{
a(t) = H^{-1} \cosh Ht  \;\;\; ; \;\;\; \Phi = {6H^2\over V_0}
}

\noindent$\bullet$ $k = -1$: There are again two solution segments, since
\interY\ requires $Y^2 \geq 1$. The cosmological solutions are given by
\eqn\mds{
a(t) = H^{-1} \sinh H|t|  \;\;\; ; \;\;\; \Phi = {6H^2\over V_0}
}
Note that these are the de-Sitter family of solutions.

\noindent{\it (iii) Attractor solutions.}

Finally, it is apparent from figure 1 that there are three main
attractors: 

\noindent ${\cal A}_0 =\{X=0,Y\to\infty\}$ 

\noindent ${\cal A}_\pm =\{-Y\propto\pm X\to \infty\}$

We have already discussed the attractor ${\cal A}_0$ which
corresponds to the de-Sitter family of universes, therefore it only
remains to discuss ${\cal A}_\pm$. Clearly, from \xyofz\ $Z\to 0$
along each of these trajectories, and to leading order we find:
\eqn\apmxyz{
\eqalign{
Y &\propto {1\over 2\eta} \; , \;\;\; Z \propto -\eta \;,
\;\;\; {\rm with\ } \eta \to 0^- \cr
a(t) &\propto |t|^{2A \pm 1\over 6A \pm 1} \;
, \;\;\; \Phi \propto |t|^{ \mp 2 \over 6A \pm 1} \; , \;{\rm with\ }
\cases{ t\to 0^- & for $6A \pm 1 > 0$. \cr
t\to\infty & for $6A - 1 <0$. \cr} \cr
a(t) &\propto e^{Ht} \; , \;\;\; \Phi \propto e^{-3Ht}\;,\;\;\; {\rm with\ }
t\to\infty \;\; {\rm for\ } 6A=1\cr
}}
Thus the attractor ${\cal A}_+$ always corresponds to a (power law)
contracting universe with a diverging JBD scalar. The ${\cal A}_-$ attractor
has a vanishing JBD scalar and is contracting for $A>1/2$, expanding for
$A<1/2$. This behaviour can be seen clearly from considering the
$a'=0$ line $Y=X/2A$. For $A>1/2$, this line lies in
the physically inaccessible $\rho<0$ regions for $k = 0,-1$, only briefly
crossing the $\rho>0$ region for $k=1$ giving rise to a small number of
bounce solutions in that case. For $A<1/2$, the $a'=0$ line lies
(almost) entirely in the allowed $\rho>0$ region, with a small exception
now for $k = -1$ where it crosses the $\rho<0$ barrier. This means
that there are many bounce solutions for $A<1/2$. In particular, 
the vacuum attractor ${\cal A}_-$ now becomes a late time expanding solution.

We are now in a position to state the general behaviour of a vacuum cosmology.
Apart from a few special $k = 1$ trajectories which cross the $X$-axis, the 
late time behaviour of a vacuum universe is determined by which
quadrant of the $(X,Y)$ plane it lies in. Trajectories with $Y>0$ attract
to ${\cal A}_0$ - the de-Sitter universes, and those with $Y<0$ attract
to ${\cal A}_\pm$ according to whether $X=\pm |X|$. The exceptions to this rule
are a small set of $k = 1$ cosmologies which correspond to trajectories 
crossing the $X$-axis. These correspond to $D<0$ trajectories and either have
$Z\geq Z_3$ or $0<Z\leq Z_2$. The former trajectories lie very close to the 
$Y$-axis and appear to be perturbations of de-Sitter universes with a
very nearly constant dilaton. The latter trajectories are perturbations
of the ${\cal A}_\pm$ attractors either having a very small dilaton
(${\cal A}_-$) or a very large dilaton (${\cal A}_+$).
We should also point out that in addition to these bouncing cosmologies, 
for $k=1$ there are also a handful of
vacillating and coasting cosmologies as indicated in the closeup
figure 1d. A vacillating cosmology has two zeros of $a'$ and corresponds
to an expanding universe which slows down, recontracts for a short
period, then re-expands (or vice versa). The coasting cosmology is
tangent to the $a'=0$ line, and represents a universe which expands, slows
down and halts expansion, then re-expands again (and vice versa). 

\subsec{The radiation cosmologies.}

The classification of the radiation cosmologies is extremely
similar to the vacuum cosmologies, as can be seen from the 
$(X,Y)$ phase plots shown in figure 4, however, there are a few
subtle differences that should be highlighted. 
\ifig\figfour{The $(X,Y)$ phase plane diagrams for the radiation
cosmologies. The shaded areas represent the disallowed regions where
$Z<0$. These figures
are for $k = 1$, the general figure for $k = 0,-1$ taking the
appearance of figure 4c. Figure 4d shows the trajectories for different
spatial curvatures with $B$ chosen to give the same disallowed region;
the solid trajectories are those with $k = 1$, the dashed, $k = 0$, and
the dotted, $k = -1$.}
{\epsfxsize 14truecm \epsfbox{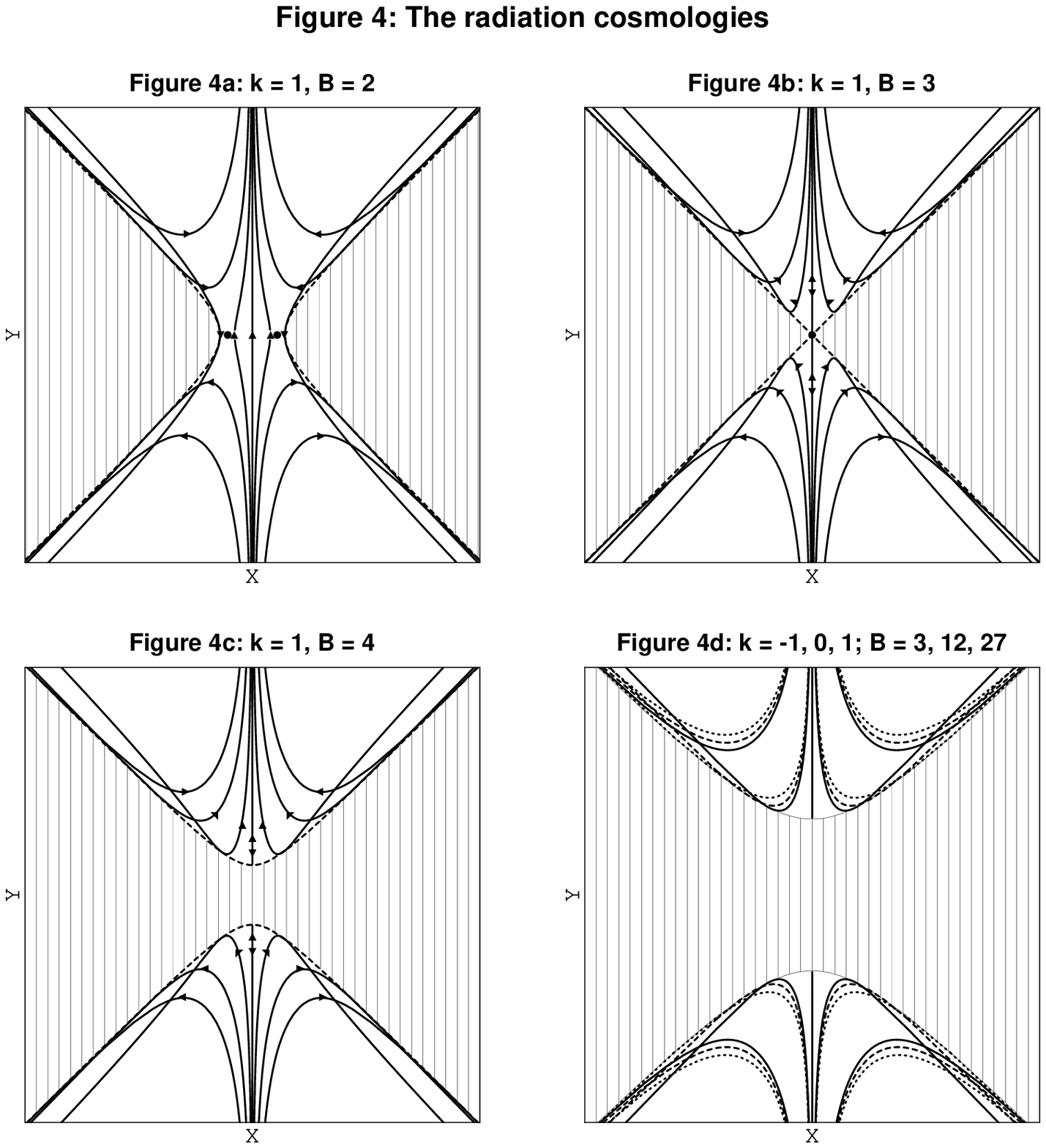}}

As we have already mentioned, the radiation $(X,Y)$ plot is a projection,
and therefore contains several features that are a consequence of this.
Most obvious  is the apparent crossing of trajectories. Recall that
different trajectories correspond to different choices of $C$
in \xyofz, therefore, when $Z_1 = \sqrt{C_1\over C_2} Z_2$ the value
of $X$ at $Z_1$ and $Z_2$ is the same. If in addition $Z_1 = 6B/Z_2$,
then the $Y$ values are the same. Since $Z$ ranges from zero to infinity
on those trajectories which have a minimum non-zero value of $|Y|$,
this means that any one such trajectory will intersect any other
such trajectory once. One implication of this is that the boundary
of the physically allowed region is no longer a trajectory
(or projective trajectory) as it was in the vacuum case. The physically
allowed region is given from \vrcon:
\eqn\rewrite{
Y^2+k-X^2 = {Z\over6} + {B\over2Z}
}
In vacuum, $Z\geq0$ gave $Y^2+k \geq X^2$, hence
the shaded regions in figure 3. If, however, $B\neq0$, the right 
hand side of \rewrite\ now has a minimum value $\sqrt{B\over3}$ for
$Z=\sqrt{3B}$, i.e.
\eqn\posdbar{
Y^2+k-X^2 \geq \sqrt{B\over3}
}
as shown in figure 4.
Therefore, unless we are at the critical point $P_0$ for $B=3$,
$Z$ cannot remain at $\sqrt{3B}$ along a trajectory. This means that
each trajectory reaches the minimum value at most once, and for
$Z\in(0,\infty)$ trajectories, at least once. Thus the curve given by
equality in \posdbar\
forms an envelope for the physical trajectories.

Another straightforward difference is that for $k = 0,-1$, there can
be no critical points, since \vrcon\ implies $|Y|,Z>0$ strictly, and
hence $|Z'|$ is strictly positive. Thus the intersection of the 
forbidden region with the $Y$-axis at $(0, ({B\over3})^{1\over4})$
does not represent a critical point, merely a stationary point on the
trajectory $Y(Z) = \sqrt{{Z\over6} + {B\over 2Z}}$. For $k=1$,
we must have $B\leq3$ for a critical point, in which case the
points correspond to $P_+$ as defined in \radcrit\ for the $(Z,W)$-plane, and
occur at $X = \sqrt{C_0}/Z_+$, where $Z_+$ is as defined in \rproot.
The critical points are  classified in table 2.
\midinsert
\begintable
\multispan{4}\tstrut\hfil {\bf TABLE 2:
The critical points of a radiation cosmology.}\hfil\crthick
\tstrut {~} | $ k = -1$ | $k = 0$ | $k = +1$ \crthick
$P_+$ | --- | --- | $({\sqrt{C_0}\over 2+\sqrt{4-B}},0)$, $B<3$, SADDLE  \cr
$P_{0,-}$ | \multispan{3} VACUUM ONLY  \endtable
\endinsert

Finally, note that the `attractors' ${\cal A}_\pm$ are subtly different
for the radiation case. Although to leading order ${\cal A}_\pm$ 
and the solutions are the same, to sub-leading order, 
\eqn\sublead{
|Y| \sim |X| + {B\over4\sqrt{C}}
}

Therefore, qualitatively the overall behaviour of the radiation cosmologies
is the same as the vacuum cosmologies. Universes starting at $Y>0$
generically attract to de-Sitter universes at late times, and those 
with $Y<0$ to the ${\cal A}_\pm$ solutions. For $k = 1$ and sufficiently
low radiation density ($B<3$) there will be a small number of 
trajectories crossing the axis, and hence universes which are a perturbation
of the full de-Sitter/ ${\cal A}_\pm$ universes, however, as we consider
initial values $|Y| \to \infty$, these form a smaller and smaller
set, tending to measure zero.
In addition, there will also be a small number of coasting and
vacillating cosmologies.

\subsec{The Einstein frame.}

Finally, we would like to remark on the naturalness of the 
de-Sitter solutions within this family by making a conformal
transformation to the Einstein frame. To do this, we set
\eqn\eframe{
g_{ab} = \Phi^{-1} {\hat g}_{ab}
}
so that
\eqn\eterms{
\sqrt{-g} \Phi R = \sqrt{-{\hat g}} {\hat R} + 3 {\hat \bo} \ln \Phi
- {3\over2} ({\hat \nabla} \ln \Phi)^2
}
and the gravitational part of the action appears in Einstein form.

Under such a transformation, the radiation energy-momentum tensor is
conformally invariant, and the potential transforms as
\eqn\pottrans{
\sqrt{-g} V(\Phi) = \sqrt{-{\hat g}} V_0
}
i.e.\ the potential appears as a cosmological constant in the
Einstein frame. Thus, for $\Phi$ constant, we would expect 
de-Sitter solutions.

\newsec{Dust and false vacuum cosmologies.}
 
In this section we analyse the dynamical system \ds\ for the 
cases of dust ($\gamma=1$) and false vacuum ($\gamma=0$). As before,
we identify the critical points of the dynamical system, 
classifying them,  and identify the
physical regions of parameter space finding sample trajectories by
numerical integration. Finally, we re-interpret these solutions as
cosmologies, in terms of the original cosmological parameters
$\Phi(t)$ and $a(t)$.
 
\subsec{The dust cosmologies.}
 
For dust, $\gamma=1$, and the dynamical system is
\eqn\dustds{
\eqalign{
X' &= -2XY + {1\over 4A} (Y^2 + k - X^2
- {Z\over6}) \cr
Y' &= -{\half} (Y^2 + k)- {3X^2\over2} + {Z\over 4} \cr
Z' &= 2ZY \cr
}}
%
\ifig\figfive{The three-dimensional phase diagram for a $k=0$
dust cosmology with $A=1$. The ${\cal A}_0$ attractor is
clearly visible.}
{\epsfxsize 14truecm \epsfbox{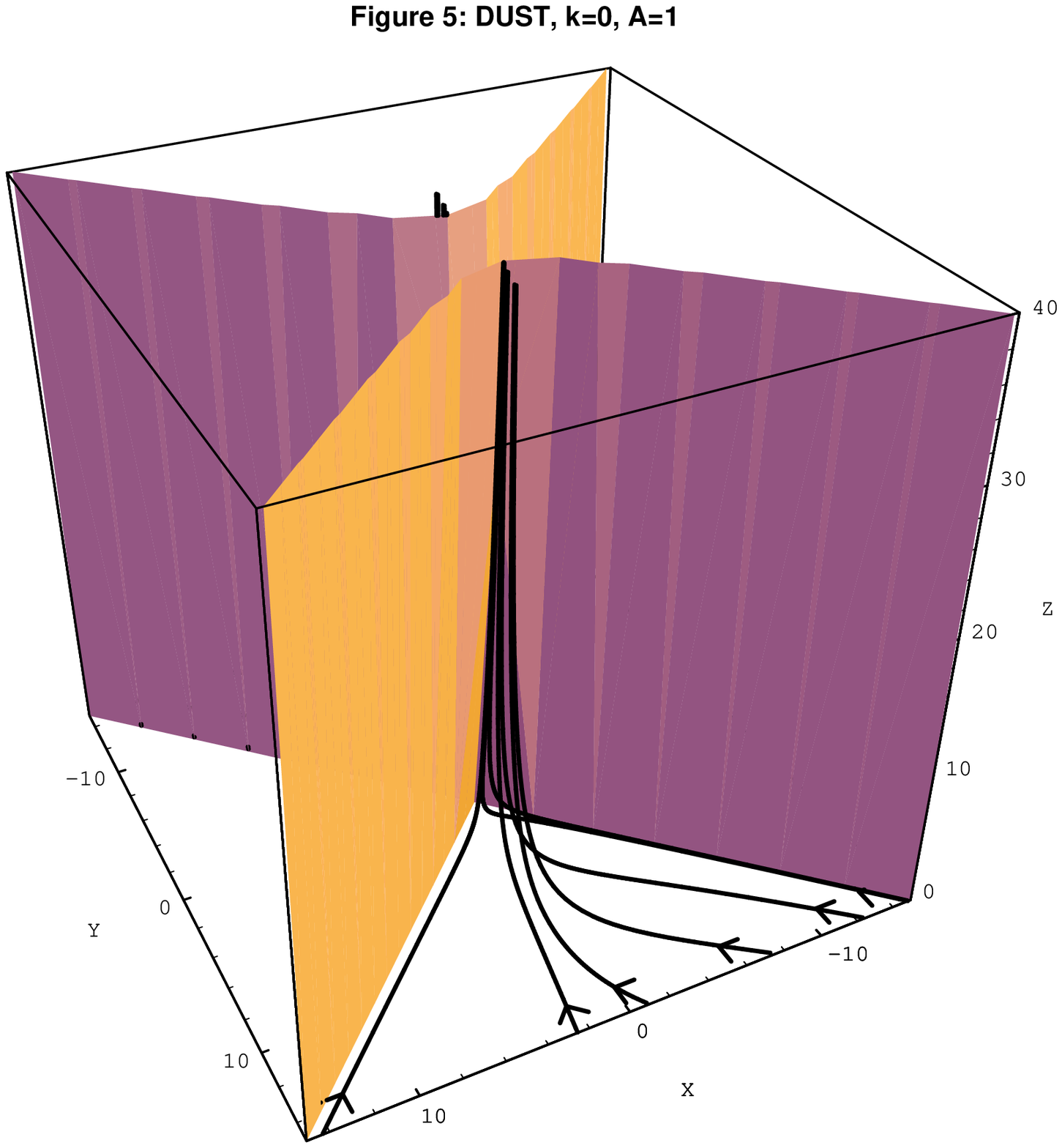}}
This apparently has six possible critical points:
\eqn\dustcrit{
\eqalign{
Q_{0,1} &= (0,\pm\sqrt{-k},0) \;\; \leftrightarrow\;\; S_{0,1} \cr
Q_{2,3} &= (\pm \sqrt{k\over 3},0,4k)\;\; \leftrightarrow\;\; S_{2,3} \cr
Q_{4,5} &= (\mp 2A \sqrt{-k\over 1+12A^2}, \pm \sqrt{-k\over 1+12A^2},0) \cr
}}
however, substitution of $Q_{4,5}$ into \interY\ indicates that it
is a non-physical critical point, as it corresponds to $\rho<0$.
For $k=0$, these are all coincident at the origin; for $k=-1$, only
$Q_{0,1}$ are relevant, and for $k=1$, only $Q_{2,3}$. Table 3 summarizes the
information concerning the critical points for the dust filled universe.
\midinsert
\begintable
\multispan{4}\tstrut\hfil {\bf TABLE 3:
The critical points of a dust cosmology.}\hfil\crthick
\tstrut {~} | $k = -1$ | $k = 0$ | $k = +1$ \crthick
$Q_{0,1}$ | $(0,\pm1,0)$ | $(0,0,0)$ | --- \crnorule
{~} | SADDLE |  DEGENERATE | ~~ \cr
$Q_{2,3}$ | --- | AS ABOVE | $({\pm1\over\sqrt{3}},0,4)$ \nr
{~}| ~ | ~ | SADDLE \cr
$Q_{4,5}$ | NOT PHYSICAL |
AS ABOVE | --- \endtable
\endinsert
Since the only physical critical points correspond to the vacuum
critical points $S_{0-3}$, we can refer to section 3.1 for the
solutions to which these correspond, namely, Minkowski spacetime
or the Milne universe for $S_{0,1}$ and \vcritneg\ for $S_{2,3}$.

Clearly, since the dynamical system is now inherently three-dimensional,
a two-dimensional plot will not convey all the information, however,
Since the sign of $Z'$ is determined by the sign of $Y$, most of the
useful information can be gleaned from an $(X,Y)$ plot. For
the purposes of visualisation, we include \figfive\  which is a 
three-dimensional plot of the $A=1$, $k = 0$ dust system.
Note how the trajectories strongly attract to $Z\to\infty, X\to 0$ in the
$Y>0$ region. This corresponds to the de-Sitter attractor ${\cal A}_0$
of the vacuum system. It is not difficult to understand this, since
for small $X$, $\Phi$ is very nearly constant, and $a\to\infty$ since
$Z\to\infty$. In other words, $\rho\propto a^{-3}\to0$ and we rapidly 
approach a vacuum situation.

For $Y<0$, $Z\to0$ rapidly along trajectories, and therefore we
do not necessarily asymptote a vacuum situation. Indeed, since $Z\simeq 0$,
we expect, and observe,
that the cosmologies will take the form of the Kolitch and
Eardley dust cosmologies, whose two-dimensional dynamical system is 
obtained from ours by setting $Z=0$. 
\ifig\figsix{The $(X,Y)$-projections of the dust cosmologies.
The ${\cal A}_0$ attractor is evident in all of these. 
Figure 6a corresponds to the projection of figure 5. The grey line
corresponds to $a'=0$ in each case.}
{\epsfxsize 14truecm \epsfbox{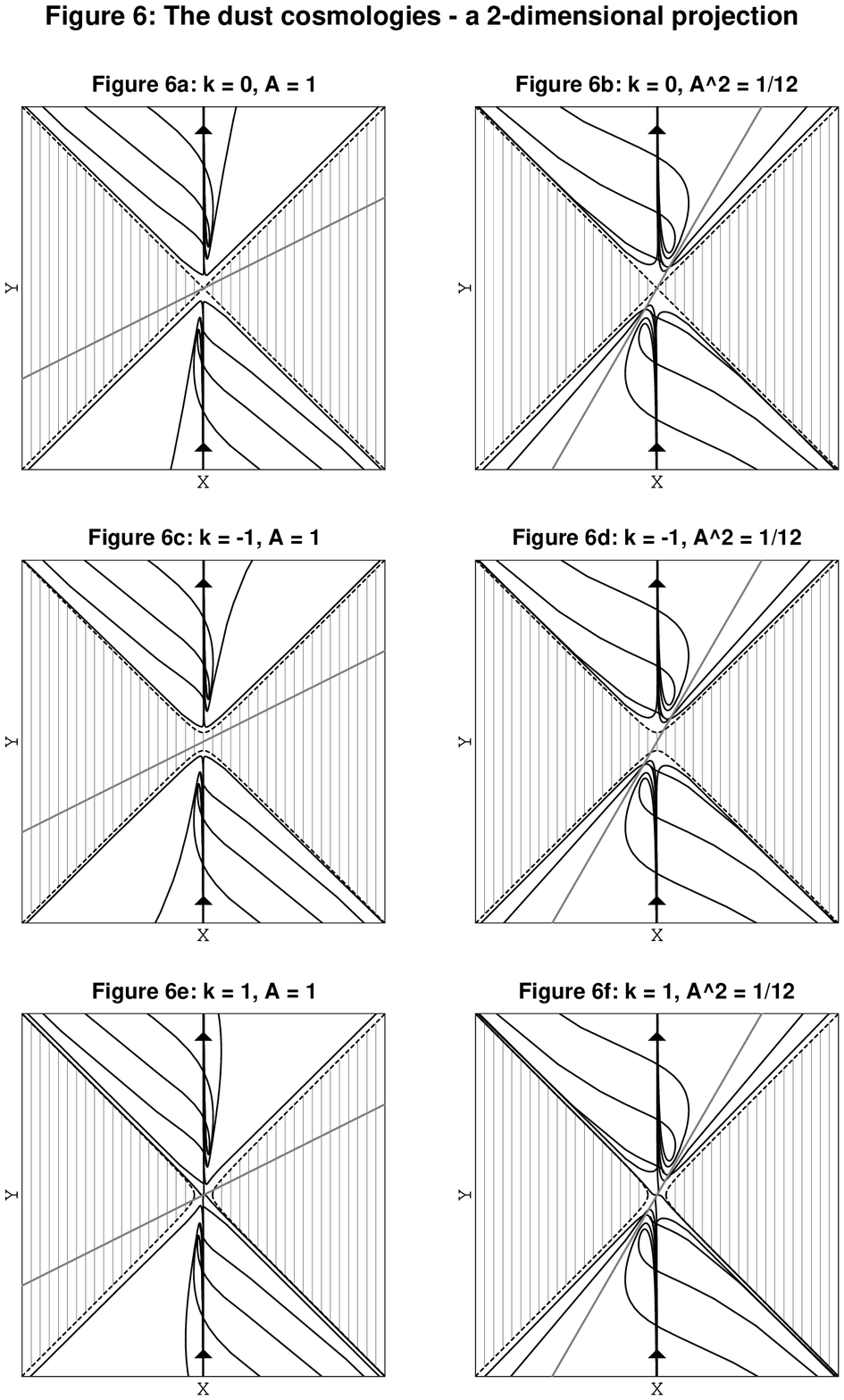}}

Figure 6 shows a selection
of two-dimensional projections of the phase space onto the $(X,Y)$ plane.
Note that ${\cal A}_+$ is weakly attractive where the divergence of $\Phi$
mitigates the divergence of $\rho$ in \interY\ to allow the cosmology
to more closely mimic the vacuum scenario.

To summarize, dust cosmologies which have $Y>0$ generally attract to de-Sitter
cosmologies. If $Y<0$ then ${\cal A}_+$ appears to weakly attract local
trajectories, but other wise there is no apparent simple classification
of the late time behaviour as in the radiation-vacuum cases. For $k=1$,
as with the radiation/vacuum cosmologies, there exist trajectories
crossing the $X$-axis. These correspond to bouncing cosmologies and take one
of two forms. If $Z\geq Z_{\rm min}$ then the trajectory stays close to
the $Y$-axis and is a perturbation of the de-Sitter universe. If
$0\leq Z \leq Z_{\rm max}$ then the trajectory start out from 
$(-\infty,\infty,0)$ and either asymptotes ${\cal A}_+$ or ${\cal A}_-$, the
bulk of such trajectories taking the former course.

\subsec{The false vacuum cosmologies.}
 
For false vacuum, $\gamma = 0$, and the dynamical system is
\eqn\fvds{
\eqalign{
X' &= -2XY + {1\over A}(Y^2 + k - X^2
- {Z\over6}) \cr
Y' &= Y^2 + k - 3X^2 \cr
Z' &= 2ZY \cr
}}
This has possible critical points
\eqn\fvcrit{
\eqalign{
R_{0,1} &= (0,\pm\sqrt{-k},0) \;\; \leftrightarrow\;\; S_{0,1} \cr
R_{2,3} &= (\pm \sqrt{k\over 3},0,4k)\;\;\leftrightarrow\;\; S_{2,3} \cr
R_{4,5} &= (\pm A \sqrt{-k\over 1-3A^2}, \pm\sqrt{-k\over 1-3A^2},0) \cr
}}
These are classified in table 4.
\midinsert
\begintable
\multispan{4}\tstrut\hfil {\bf TABLE 4:
The critical points of a false vacuum cosmology.}\hfil\crthick
\tstrut {~} | $k = -1$ | $k = 0$ | $k = +1$ \crthick
$R_{0,1}$ | $(0,\pm,0)$ | (0,0,0)| ---  \crnorule
{~} | SADDLE | DEGENERATE | \cr
$R_{2,3}$ | --- | AS ABOVE | $({\pm1\over \sqrt{3}},0,4)$ \crnorule
{~} | \ | | SADDLE \cr
$R_{4,5}$ | $\pm \sqrt{1\over1-3A^2}(A,1,0)\;;\; A<1/\sqrt{3}$ 
| AS ABOVE |
$\pm \sqrt{-1\over{1-3A^2}}(A,1,0) \;;\; A>1/\sqrt{3}$ \crnorule
{~}| $A\in(0,1/2)$, FOCUS ||SADDLE \crnorule
{~}| $A\in(1/2,1/\sqrt{3})$, SADDLE  | | \endtable
\endinsert
%
\ifig\figseven{The $(X,Y)$-projections for the false vacuum cosmologies.
Note the change in the attractors for $Y<0$ for the different 
values of $A$. The grey line
corresponds to $a'=0$ in each case.}
{\epsfxsize 14truecm \epsfbox{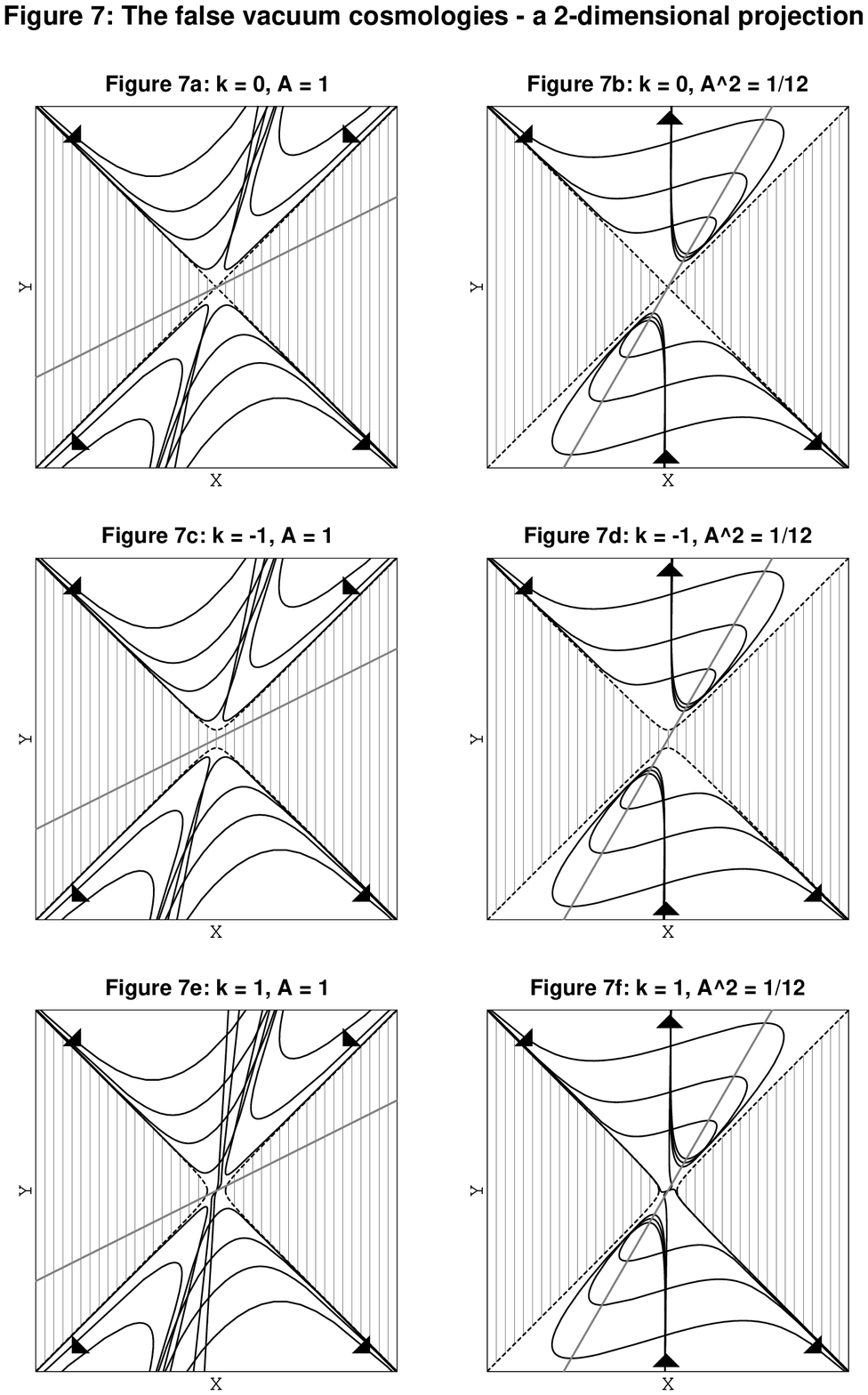}}

As before, we can refer to the vacuum critical point solutions for $R_{0-3}$,
what is less clear is that $R_{4,5}$ are also essentially the same
cosmological solutions. To see this note that $Z\equiv0$ requires
either $\Phi=0$ or $a=0$, \interY\ then implies $\Lambda=0$ or
$\infty$ respectively -- a somewhat artificial limit. Setting $\Phi = 0$
in order to get a meaningful cosmology then gives $a \propto |t|$. This
is, in each case, a coordinate transform of Minkowski spacetime and is
equivalent to the $P_0$ critical point of the $(Z,W)$-plane.

As with the dust cosmologies, the false vacuum dynamical system is also
three-dimensional, and as before, we plot the two-dimensional
projection onto the $(X,Y)$-plane in figure 7. The false vacuum case
however, seems to be the opposite of dust, since for $Y>0$ there is no
obvious attractor, while for $Y<0$ the ${\cal A}_\pm$ attractors show
up clearly. We have checked this behaviour by compactifying to the 
Poincar\'e sphere and analysing the critical points
at infinity which confirms that there is no attractor for $Y>0$.
Again, this is not difficult to understand, since the only
difference between the false vacuum dynamical system \fvds\ and the vacuum
one \trvacsis\ is the presence of the extra term in the $X'$ equation, 
which is proportional to $\Lambda a^2/\Phi$. But $(\ln{a^2\over\Phi})' \propto
(Y-X/A)$ hence  if $Y<X/A$, as it is for both ${\cal A}_\pm$ for $A\geq1$
or for ${\cal A}_+$ for all $A$, then this term is rapidly damped and we
approach the vacuum situation. For $Y>0$ however, this extra term in $X'$
grows and we depart increasingly from vacuum at late times. Thus, in 
distinction to all the other cosmologies we have considered, the 
expanding false vacuum cosmologies do not attract to de-Sitter universes at
late times.

\newsec{\bf Conclusions}

In this paper we have presented a dynamical systems analysis of the
FRW cosmologies in Brans-Dicke theory with a potential for the scalar
JBD field. Our analysis was based on that of Kolitch and others, \KE,
who found that cosmology in pure JBD theory could be expressed as
a two-dimensional dynamical system. In the presence of a scalar
potential, we found that unless the matter source was radiation, the
dynamical system was inherently three-dimensional. In spite of this
we were able to see clearly that the asymptotic late time behaviour
of the expanding cosmologies was generically that of a de-Sitter 
universe, the specifics 
depending on the topology of the spatial sections of the cosmology,
and the only exception being that of a false-vacuum source.
Since the potential chosen is actually equivalent to an Einstein
cosmological constant, these results are in general agreement with
the statement that general relativity is an attractor for Brans-Dicke
theory\DaN, although this statement comes with certain caveats, as the
false vacuum case illustrates.

In order to get an idea of when 
general relativity is a late time attractor, consider
the intermediate equation for $X'$, \interX, substituting the 
behaviour of $\rho$ from \rhoofa:
\eqn\explore{
X' = -2XY + (1-3\gamma/4) {8\pi \rho_0\over 3A} a^{2-3\gamma} \Phi^{-1}
}
Now, 
$$(\ln[a^{2-3\gamma} \Phi^{-1}])'\propto (2-3\gamma)Y-3(1-\gamma)X/A$$
therefore, if $\gamma>2/3$,  $[a^{2-3\gamma} \Phi^{-1}]$ will decrease
along a positive $Y$ trajectory, and $X$ will be driven to zero.
$\Phi$ will therefore become constant and we will asymptote a de-Sitter
universe. For $\gamma<2/3$ however, we do not expect a late
time de-Sitter solution. Since $\rho + 3p = (3\gamma-2)\rho$, $\gamma<2/3$
corresponds to a matter source which does not satisfy the strong
energy condition, therefore it appears that general relativity is
an attractor for matter obeying the SEC.

Clearly this model is somewhat artificial since it neither
anchors $\Phi$ at a specific value, nor behaves as we might
expect a scalar potential to behave, however, it does illustrate differences
that crop up between pure JBD theory and those with a scalar potential.
In particular the appearance of a strong attractor for expanding universes.
However the model also displays many similarities with the pure JBD case, 
for instance, the presence of vacillating or coasting cosmologies for $k=1$.

Perhaps the most important conclusion to draw from this work is that
for ``ordinary matter'' (i.e.\ matter obeying the strong energy condition) 
general relativity is an attractor, therefore, independent of the
specifics of a dilaton potential, i.e.\ how and at what energy scale the
dilaton acquires a mass and gravity becomes Einstein in nature, the
universe should exit the stringy era of gravity in a reasonably familiar
Einstein form. On the other hand, at higher energies, when an
inflationary phase may have occurred, there is no strong attractor(s)
-- very much in contrast to pure JBD \KE\ -- hence (as was hinted
in \HKVW) the behaviour of
inflationary universes in this case could be strongly affected by the
behaviour of the dilaton potential.

\centerline{\bf Acknowledgements}
 
We would like to thank Bob Johnson and Filipe Bonjour for conversations
and assistance with numerical aspects of this problem. C.S.\ would like
to thank the Theoretical Physics Group of the University of Oporto, in
particular Prof.E.S.Lage and Dr.J.C.Mour\~ao (IST-Lisbon)
for help in the early stages of this project.
This work was supported by a JNICT fellowship BD/5814/95 (C.S.), and
a Royal Society University Research Fellowship (R.G.).

\listrefs

\bye